\documentstyle[12pt]{article}

\newcommand{\be}{\begin{equation}}
\newcommand{\ee}{\end{equation}}
\newcommand{\bea}{\begin{eqnarray}}
\newcommand{\eea}{\end{eqnarray}}

\def\lm{\lambda}

\def\dely{\partial_y}

\newcommand\hci{H^c_i}

\newcommand\hi{H_i}

\def\x5{x_5}

\begin{document}
\setlength{\baselineskip}{0.7cm}

\begin{titlepage}
\null
\begin{flushright}
hep-ph/0306229 \\
June, 2003
\end{flushright}
\vskip 1cm
\begin{center}
{\Large\bf 
Fat Branes, Orbifolds and Doublet-Triplet Splitting
} 

\lineskip .75em
\vskip 1.5cm

\normalsize

{\large Naoyuki Haba$^{a}$} 
and {\large Nobuhito Maru$^{b}$}
\footnote{Special Postdoctoral Researcher.}

\vspace{15mm}

{\it $^{a}$Institute of Theoretical Physics, \\
 University of Tokushima, Tokushima 770-8502, JAPAN} \\
\vspace*{1cm}
{\it $^{b}$Theoretical Physics Laboratory, \\
RIKEN (The Institute of Physical and Chemical Research), \\
2-1 Hirosawa, Wako,  Saitama 351-0198, JAPAN} \\

\vspace{18mm}

{\bf Abstract}\\[5mm]
{\parbox{13cm}{\hspace{5mm}
%
%%%%%%%%%%%%%%%%%%%%%%%%%%%%%%%%%%%%%%%%%%%%%%%%%%%%%%%%%%%%%%%%%%
A simple higher dimensional mechanism of the doublet-triplet 
 splitting is presented in a five dimensional supersymmetric SU(5) 
 GUT on $S^1/Z_2$. 
The splitting of multiplets is realized 
 by a mass term of Higgs hypermultiplet 
 which explicitly breaks SU(5) gauge symmetry. 
Depending on the sign of mass, 
 zero mode Higgs doublets and triplets are localized on 
 the either side of the fixed points. 
The mass splitting is realized due to the difference of 
 magnitudes of the overlap with a brane localized 
 or a bulk singlet field. 
No unnatural fine-tuning of parameters is needed. 
The proton stability is ensured by locality {\em without symmetries}. 
As well as a conventional mass splitting solution, 
 it is shown that the weak scale Higgs triplet 
 is consistent with the proton stability. 
This result might provide an alternative signature of GUT 
 in future collider experiments. 
%%%%%%%%%%%%%%%%%%%%%%%%%%%%%%%%%%%%%%%%%%%%%%%%%%%%%%%%%%%%%%%%%%%
}}

\end{center}

\end{titlepage}

The doublet-triplet splitting problem is one of the notorious 
 problems in grand unified theories (GUTs) \cite{GUT}. 
Many proposals for this problem have been made in four dimensional models 
\cite{4dDT} or recently in higher dimensional models \cite{hdDT}. 
See also for string-derived models \cite{string}. 
In our previous papers \cite{Maru,HabaMaru}, 
 we have presented a doublet-triplet splitting mechanism in the context 
 of the fat brane scenario \cite{AS}. 
In this approach, a conventional mass splitting, 
 namely the weak-GUT scale splitting, can be realized 
 by an overlap of zero mode wavefunctions between the doublet and 
 the triplet Higgs fields without unnatural fine-tuning of parameters 
 \cite{Maru}. 
Interestingly, it has also been shown that 
 an alternative mass splitting, namely the weak-TeV scale 
 splitting, can be realized by the same machanism \cite{HabaMaru}. 
The proton stability is guranteed by the strong suppression of 
 the coupling of matter fields to the triplet Higgs 
 due to a small overlap of wavefunctions. 
This TeV scale triplet Higgs scenario deserves an attention as 
 an alternative signature of GUT instead of the proton decay.

The setup itself has crucial problems 
 although these scenarios are very attractive. 
In the fat brane scenario, 
 extra dimensions are considered to be non-compact 
 otherwise the effective four dimensional theory becomes vector-like. 
This non-compactness leads that the gravity and gauge fields 
 cannot propagate in the bulk and have to be localized on a fat brane 
 (domain wall) to obtain finite coupling constants. 
As is well known, it is highly nontrivial to realize the localization 
 of the gravity and gauge fields on a domain wall 
 in infinite extra dimensions. 
Even if the localization is realized by the ``quasi-localization" mechanism 
 \cite{DGS,Kolanovic}, 
 zero mode wave functions have to be almost flat on a domain wall 
 because of the charge universality constraints. 
It seems to be very difficult to obtain such flat wave functions.

To avoid this situation, 
 we consider a theory on an orbifold.\footnote{In the smooth 
 compact dimensional case, $S^1$ for instance, 
 the anti-domain wall as well as the domain wall appears. 
 This case cannot only yield chiral fermions but also cause 
 the instability of the system.} 
By construction, the bulk gravity and gauge fields propagate 
 in a finite extra space and it is easy to obtain flat zero mode 
 wave functions in flat extra dimensions. 
First explicit realization of the fat brane scenario on an orbifold 
 has been given in Ref.~\cite{GGH}. 
By developing extra coordinate dependent vacuum expectation 
 values (VEVs) of the parity odd scalar field, 
 the kink solution (domain wall) is constructed. 
The chiral fermion zero modes which couples to the scalar field 
 generating domain wall are localized on either fixed point 
 depending on the sign of the coupling constant.

In this letter, 
 we apply this mechanism to the doublet-triplet splitting 
 in a five dimensional supersymmetruc (SUSY) SU(5) GUT 
 compactified on $S^1/Z_2$.\footnote{For other applications, 
 see \cite{KT,GP}.} 
There are fixed points at $y=0,\pi R$, 
 where $y$ denotes the fifth dimensional coordinate and 
 $R$ is a compactification radius.

We shall focus on a Higgs sector. 
Two hypermultiplets are introduced as Higgs multiplets. 
\bea
H_1 =(H_1({\bf 5}), H_1^c({\bf 5^*})),
\quad H_2 =(H_2({\bf 5^*}), H_2^c({\bf 5})), 
\eea
where the representations under SU(5) are specified. 
We assign their $Z_2$ parities as follows;
\bea
&&H_1(-y) = + H_1(y), \quad H_1^c(-y) = - H_1^c(y), \\
&&H_2(-y) = + H_2(y), \quad H_2^c(-y) = - H_2^c(y). 
\eea
The action we consider is given by
\bea
\label{action}
S_5 &=& \int d^5 x 
\left\{
[\hi^\dag e^{-V} \hi + \hci e^V H^{c\dag}_i]_{\theta^2 \bar{\theta}^2} 
+ [\hci(\dely - \frac{1}{\sqrt{2}}\Phi)\hi \right. \nonumber \\
&& \left. - \frac{1}{2} M(y) H_i^c H_i 
+ \delta(y - \pi R) \frac{S}{M_P} H_1 H_2]_{\theta^2} + {\rm h.c.}
\right\}~(i=1,2), 
\eea
where the action $S_5$ is written in terms of 4D, ${\cal N}=1$ superspace 
formalism \cite{4dsuperspace}, 
$S$ is a singlet chiral superfield, 
 which is assume to be localized on the brane at $y=\pi R$.\footnote{
 The cases that is localized at $y=0$ and constant in the bulk 
 will be discussed later.}  
$M_5,M_P$ are Planck scales in five and four dimensions, 
which are related by $M_5^3 \pi R = M_P^2$. 
Here, the vector superfield $V$ and the chiral superfield $\Phi$
in the adjoint representation are explicitly given by 
\bea
V &=& -\theta \sigma^\mu \bar{\theta} A_\mu + i\bar{\theta}^2 \theta \lm_1 
-i\theta^2 \bar{\theta} \bar{\lm}_1 + \frac{1}{2}\theta^2 \bar{\theta}^2 D, \\
\Phi &=& \frac{1}{\sqrt{2}}(\Sigma + iA_5) + \sqrt{2} \theta \lm_2 + \theta^2 F, 
\eea
where $A_\mu~(\mu =0,1,2,3)$ is a gauge field in four dimensions, 
 $\lambda_{1,2}$ are gauginos, 
 $F, D$ are auxiliary fields, 
 $A_5$ is an extra component of the gauge field, 
 $\Sigma$ is a real scalar field in the adjoint representaion. 
Note that an extra dimensional coordinate dependent mass term 
is introduced in the second line of Eq.~(\ref{action}). 
A mass parameter $M(y)$ is given by 
\bea
\label{hypermass}
M(y) = M_5~{\rm diag}(2,2,2,-3,-3) \varepsilon(y), 
\eea
where $\varepsilon(y)$ is a sign function with respect to $y$. 
Thus, an SU(5) is explicitly broken to the standard model (SM) gauge group 
 at the cutoff scale $M_5$. 
%{\bf Comments on unitarity should be added.}
%Furthermore, this hypermultiplet is assumed to be a spurion field, 
%in other words, this hypermultiplet has no kinetic term to be consistent with 
%SUSY in five dimensions and only scalar component has a VEV. 
Throughout this paper, 
we consider the following SUSY vacuum at $M_5$\footnote{Later, 
we consider the VEV of $S$. 
We assume that $S$ will take VEV below 
the energy scale of $M_5$ by, for an example, 
an inverted hierarchy scenario.}
\bea
\label{vacuum}
\langle \Sigma \rangle = \langle H_i \rangle = \langle H^c_i \rangle =
\langle S \rangle = 0. 
\eea

Let us concentrate on fermionic components of the action 
to study zero modes in the background (\ref{hypermass}) and (\ref{vacuum}), 
\bea
S_5 &\supset& 
\psi_i^c i \sigma^\mu \partial_\mu \bar{\psi}_i^c 
+ \bar{\psi}_i i \bar{\sigma}^\mu \partial_\mu \psi_i 
- \psi_i^c \partial_y \psi_i + \bar{\psi}_i \partial_y \bar{\psi}_i^c 
+ \frac{1}{2} M(y) 
(\psi_i^c \psi_i + \bar{\psi}_i^c \bar{\psi}_i) \nonumber \\
&&-\delta(y-\pi R)(\frac{S}{M_P} \psi_1 \psi_2 
+ \frac{S^*}{M_P} \bar{\psi}_1 \bar{\psi}_2)~(i=1,2), 
\eea
%where $\Psi^T_5 =(\psi, \bar{\psi}^c)$ is a Dirac fermion composed 
%of 2-component fermions $\psi$ and $\bar{\psi}^c$. 
where $(\psi_i, \psi_i^c)$ denote obviously the fermionic components 
of $(H_i, H_i^c)$. 
By expanding in modes as 
\bea
\psi_i(x,y) = \sum_n \psi_i^{(n)}(x) f^{(n)}(y),\quad 
\psi_i^c(x,y) = \sum_n \psi_i^{c(n)}(x) f^{c(n)}(y), 
\eea
we obtain mode equations 
\bea
\label{m1'}
0 &=& m_n \bar{\psi}_1^{(n)} f_1^{c(n)} 
+ \bar{\psi}_1^{(n)} \partial_y \bar{f}_1^{(n)} 
- \frac{1}{2} M(y) \bar{\psi}_1^{(n)} \bar{f}_1^{(n)}, \\
\label{m2'}
0 &=& m_n \bar{\psi}_2^{(n)} f_2^{c(n)} 
+ \bar{\psi}_2^{(n)} \partial_y \bar{f}_2^{(n)} 
- \frac{1}{2} M(y) \bar{\psi}_2^{(n)} \bar{f}_2^{(n)}, \\
\label{m3'}
0 &=& m_n \bar{\psi}_1^{c(n)} f_1^{(n)} 
-\bar{\psi}_1^{c(n)} \partial_y \bar{f}_1^{c(n)} 
-\frac{1}{2} M(y) \bar{\psi}_1^{c(n)} \bar{f}_1^{c(n)} 
+ \delta(y-\pi R) \frac{S^*}{M_P} \bar{\psi}_2^{(n)} \bar{f}_2^{(n)}, \\
\label{m4'}
0 &=& m_n \bar{\psi}_2^{c(n)} f_2^{(n)} 
-\bar{\psi}_2^{c(n)} \partial_y \bar{f}_2^{c(n)} 
-\frac{1}{2} M(y) \bar{\psi}_2^{c(n)} \bar{f}_2^{c(n)} 
+ \delta(y-\pi R) \frac{S^*}{M_P} \bar{\psi}_1^{(n)} \bar{f}_1^{(n)}, 
\eea
where a mass in four dimensions is defined as
\bea
-i\sigma^\mu \partial_\mu \bar{\psi}_i^{c(n)} = m_n \psi_i^{(n)},\quad 
-i \bar{\sigma}^\mu \partial_\mu \psi_i^{c(n)} = m_n \bar{\psi}_i^{(n)}. 
\eea
It is easy to find solutions of above equations 
 (\ref{m1'})-(\ref{m4'}) in the bulk, 
\bea
f_{1,2}^{(n)}(y) &=& N_n {\rm exp}
\left[ \frac{1}{2}\int_0^y dx_5 M(x_5) \right]
{\rm cos}(m_n y), \\
f_{1,2}^{c(n)}(y) &=& N_n^c {\rm exp}
\left[ -\frac{1}{2}\int_0^y dx_5 M(x_5) \right]
{\rm sin}(m_n y), 
\eea
where $N_n^{(c)}$ are normalization constants.

On the other hand, 
 the following boundary conditions at $y=\pi R$ should be satisfied 
 from (\ref{m3'}) and (\ref{m4'}), 
%\bea
%0 &=& 2 \psi^c_1~{\rm sin}(m_n \pi R)
%~{\rm exp}\left[ -\frac{1}{2}\int_0^{\pi R} dy M(y) \right] 
%\nonumber \\
%&&+ \frac{S}{M_P} \psi_2~{\rm cos}(m_n \pi R)~{\rm exp}
%\left[ \frac{1}{2}\int_0^{\pi R} dy M(y) \right],
%\eea
%which can be simplified as 
\bea
\label{bc1}
{\rm tan}(m_n \pi R) &=& 
-\frac{S}{2M_P}\frac{\psi_2}{\psi_1^c}{\rm exp} 
\left[\int_0^{\pi R} dy M(y) \right], \\  
%\eea
%A similar condition from (\ref{m4'}) should be satisfied as 
%\bea
\label{bc2}
{\rm tan}(m_n \pi R) &=& 
-\frac{S}{2M_P}\frac{\psi_1}{\psi_2^c}{\rm exp}
\left[\int_0^{\pi R} dy M(y) \right]. 
\eea
Mass eigenvalues can be obtained by eliminating $\psi_{1,2}$ 
in (\ref{bc1}) and (\ref{bc2}) as 
\bea
m_n = \frac{1}{R}
\left(n + \frac{1}{\pi}{\rm arctan}
\left[\frac{S}{2M_P}{\rm exp} 
\left(\int_0^{\pi R} dy M(y) \right) \right]
\right)~(n=0,1,2...). 
\label{mass}
\eea
For $m_0=0$, 
%\footnote{$\langle S \rangle$ is assumed 
%to take VEV below the GUT scale.}, 
zero mode wave functions take the form
\bea
f_{1,2}^{(0)}(y) &\simeq& {\rm exp}
\left[
\frac{1}{2}\int^y_0 dx_5 M(x_5) 
\right], \\
&=& 
\left\{
\begin{array}{l}
\sqrt{ \frac{2M_5}{e^{2 M_5 \pi R}-1}} 
{\rm exp}[M_5 y]~({\rm for~triplets}) 
\\
\\
\sqrt{ \frac{3M_5}{1-e^{-3 M_5 \pi R}}} 
{\rm exp}[-\frac{3}{2} M_5 y]~({\rm for~doublets}) \\
\end{array}
\right..
\eea
It turns out that the triplet Higgs zero mode is peaked at $y=\pi R$, 
while the doublet Higgs zero mode at $y=0$.

The doublet-triplet splitting is realized by 
 the coupling of Higgs doublets to $S$ 
\bea
\label{DTsplitting}
\delta(y-\pi R)
\left[ \frac{S}{M_P} H_1 H_2 \right]_{\theta^2} 
\Rightarrow 
\left\{
\begin{array}{l}
m_3 = \frac{\langle S \rangle}{M_P} 
\frac{2M_5}{e^{2 M_5 \pi R}-1} e^{2M_5 \pi R} 
\simeq \frac{\langle S \rangle}{M_P} 2 M_5 \\
\\
m_2 = \frac{\langle S \rangle}{M_P} 
\frac{3M_5}{1-e^{-3 M_5 \pi R}} e^{-3M_5 \pi R} 
\simeq \frac{\langle S \rangle}{M_P} 3 M_5 e^{-3M_5 \pi R} 
\simeq m_W \\
\end{array}
\right.,
\eea
where the weak scale $m_W$ is $m_W \simeq 100$ GeV, 
$m_{2,3}$ are the doublet, triplet Higgs masses respectively. 
Remarkably, the triplet mass is unsuppressed since an overlap 
 with a singlet is large, while the doublet mass is {\em exponentially} 
 suppressed since an overlap with a singlet is very small.

In order to constrain parameters in our model further, 
 let us study the proton decay. 
We assume that the matter fields are localized on the brane at $y=0$. 
Then, the coupling of matter fields to the triplet Higgs $H_3$
 is given by 
\bea
\delta(y)
\left[\frac{Y}{\sqrt{M_P}} H_3 QQ \right]_{\theta^2}, 
\eea
where $Y$ is Yukawa coupling, $Q$ mean the SM matter superfields. 
%and $M_P$ is a 4D Planck scale. 
Integrating out with respect to the fifth coordinate $y$ 
 and substituting the value of zero mode wave functions of Higgs triplets 
 at $y=0$ yield a 4D effective yukawa coupling $Y_{{\rm eff}}$, 
which is determined by a normalization constant, 
\bea
\label{pdecay1}
Y_{{\rm eff}} \simeq \sqrt{\frac{2M_5}{M_P}} e^{-M_5 \pi R}. 
\eea
Let us first discuss proton decay constraints from dimension six operator.  
There are two sources for dimension six operator, 
one is X, Y gauge boson exchange, 
the other is the triplet Higgs scalar exchange at tree level. 
In our scenario, 
the amplitude of X, Y gauge boson exchange is $1/m^2_{X,Y} \simeq 1/M_5^2$, 
where $m_{X,Y}$ are X, Y gauge boson masses of order 5D Planck scale $M_5$ 
since GUT symmetry is broken by VEV of a hypermultiplet 
in the adjoint representation $\langle A^c A \rangle$. 
Namely, the constraint is simply $M_5 > 10^{16}$~GeV. 
%This means that the dominant contribution is always coming from 
%a triplet Higgs boson exchange diagram 
%since the triplet Higgs boson mass is smaller than 5D Planck scale. 
On the other hand, 
the amplitude of the proton decay process 
from the triplet Higgs boson exchange is 
\bea
\label{dim6}
\frac{(Y_{{\rm eff}})^2}{m_3^2} 
&\simeq& \left( \frac{M_P}{2 M_5 \langle S \rangle} \right)^2 
\left( \frac{2M_5}{M_P} e^{-2M_5 \pi R} \right)
= \frac{M_P}{2M_5 \langle S \rangle^2} e^{-2 M_5 \pi R}. 
\eea
Therefore, the triplet Higgs boson exchange constraints to be satisfied 
leads to
\be
\label{dim6con}
\sqrt{\frac{2 M_5}{M_P}} \langle S \rangle e^{M_5 \pi R} > 10^{16}~{\rm GeV}. 
\ee
Using the second relation in (\ref{DTsplitting}), 
the upper bound on the compactification scale is obtained as
\be
\label{ubm5}
R^{-1} < 5.5 \times 10^{17}~{\rm GeV}. 
%M_5 < 8.5 \times 10^{17}~{\rm GeV}. 
\ee

Next we turn to the constraints from dimension five operators. 
The proton decay amplitude coming from 
 the dimension five operator is evaluated 
\bea
\label{pdecay3}
\frac{g^2}{16 \pi^2} \frac{Y^2_{{\rm eff}}}{m_3 M_\lambda} 
\simeq \frac{g^2}{16 \pi^2} 
\frac{y_1 y_2 \left( 
\frac{2 M_5}{M_P} e^{-2 M_5 \pi R} \right)}{m_3 M_\lambda}, 
\eea
where $g$ is an SU(2) gauge coupling, $y_{1,2}$ are Yukawa coupling of 
the first and the second generations in five dimensions, 
and $M_\lambda$ is a gaugino mass. 
Comparing 4D GUT case, the following condition can be obtained, 
\bea
&&\frac{1}{\langle S \rangle} \times 10^{16} e^{-2 M_5 \pi R} < 1 
\Leftrightarrow 
\frac{3 M_5}{M_P} \times 10^{14} e^{-5 M_5 \pi R} < 1, 
\label{dim5con}
\eea
the second expression can be obtained by using 
the second relation in (\ref{DTsplitting}). 
Summarizing the proton decay constraints, 
X, Y gauge boson exchange amplitudes proportional to $M_5^{-2}$ 
gives a lower bound for $M_5$, namely a upper bound 
for the compactification scale $R^{-1}$. 
On the other hand, 
the amplitudes from Higgs triplet scalar exchange 
and the dimension five operator are roughly proportional to 
$M_5/M_P$ for $M_5/M_P \ll 1$, 
therefore a upper bound for $M_5$ is obtained, 
namely a lower bound for the compactification scale is obtained. 
Searching for allowed parameter region satisfying 
the constraints of X, Y gauge boson exchange, 
(\ref{ubm5}) and (\ref{dim5con}) with $M_5^3 \pi R = M_P^2$, 
 the following results are obtained
\bea
%\label{M5}
%7.0 \times 10^{17}~{\rm GeV} < &M_5& < 8.5 \times 10^{17}~{\rm GeV}, \\
\label{1/R}
1.8 \times 10^{17}~{\rm GeV} < &R^{-1}& < 5.5 \times 10^{17}~{\rm GeV}, \\ 
\label{S}
2.6 \times 10^9~{\rm GeV} < &\langle S \rangle& < 
6.7 \times 10^{17}~{\rm GeV}, \\
\label{colored}
2.1 \times 10^9~{\rm GeV} < &m_3& < 3.9 \times 10^{17}~{\rm GeV}, 
%m_2 &\simeq& 100~{\rm GeV},
\eea
where the upper (lower) bounds of (\ref{1/R}) 
((\ref{S}) and (\ref{colored})) come from 
the dimension six proton decay constraints, 
while the other bounds come from the cutoff scale 
$\langle S \rangle$ ($\langle S \rangle < M_P$). 
Note that the upper bound for $\langle S \rangle$ becomes more stringent 
since $\langle S \rangle$ depends on $M_5$ or $R^{-1}$ 
through the doublet Higgs mass in (\ref{DTsplitting}). 
A large $\langle S \rangle$ corresponds to a small $M_5$, 
namely a small $R^{-1}$ and vice versa. 
It is very interesting in that 
 an intermediate scale triplet Higgs can be consistent with 
 the proton stability. 
It should be stressed that the proton stability can be ensured 
not by symmetries but by the locality only, namely by storng suppression 
of coupling constants coming from overlap of wave functions 
between the triplet Higgs and the SM matter. 
%The fact that the compactification scale is close to 
% the GUT scale is welcome from the viewpoint of the gauge coupling unification. 

So far, we have discussed the doublet-triplet splitting and 
the proton stability in the case with localized singlet at $y=\pi R$.
It is also interesting other situations.  
What's happen when $S$ is localized at $y=0$? 
In this case, the splitting is realized by 
\bea
\label{y=0}
\delta(y) \left[ \frac{S}{M_P} H_1 H_2 \right]_{\theta^2} 
\Rightarrow 
\left\{
\begin{array}{l}
m_3 = \frac{\langle S \rangle}{M_P} 
\frac{2M_5}{e^{2 M_5 \pi R}-1} \simeq 
2M_5 \frac{\langle S \rangle}{M_P} e^{-2M_5 \pi R} \\
\\
m_2 = \frac{\langle S \rangle}{M_P} 
\frac{3M_5}{1-e^{-3 M_5 \pi R}} 
\simeq 3M_5 \frac{\langle S \rangle}{M_P} 
\simeq m_W~(\simeq 100~{\rm GeV}) \\
\end{array}
\right., 
\eea
then the ratio between the doublet and the triplet Higgs masses 
 is modified as 
\bea
\frac{m_3}{m_2} 
%\simeq \frac{2M_5 e^{-2M_5 \pi R}
%\langle S \rangle/M_P}{3M_5 \langle S \rangle/M_P} 
\simeq \frac{2}{3} e^{-2M_5 \pi R} \ll 1. 
\eea
The triplet Higgs mass is extremely smaller than the doublet Higgs mass 
 since an overlap between the triplet Higgs localized on $y=\pi R$ and 
 $S$ is very small. 
The following argument leads that this case is incosistent 
with the proton stability. 
Consider the dimension six operator constraints 
from the triplet Higgs boson exchange (\ref{dim6con}). 
The result is 
\be
\label{lbs}
\sqrt{\frac{2M_5}{M_P}} \langle S \rangle e^{-M_5 \pi R} > 10^{16}~{\rm GeV}. 
\ee
%This result and the second relation in (\ref{y=0}) leads to 
%an upper bound on the compactification scale, 
%\be
%R^{-1} < 2.8 \times 10^{-7}~{\rm eV}. 
%M_5 < 8~{\rm TeV}. 
%\ee
%This is clearly inconsistent with our observation. 
%$\langle S \rangle \gg M_5$ described in the footnote 5. 
One can easily check that (\ref{lbs}) has no solution. 
Thus, this case is exculded.

Finally, we consider the case 
 that $S$ is a bulk field and has a constant profile. 
In this case, the mass splitting is realized by both 
 (\ref{DTsplitting}) and (\ref{y=0}), which means 
\bea
\label{bulk}
m_3 \simeq 2 M_5 \frac{\langle S \rangle}{M_P},\quad
m_2 \simeq 3 M_5 \frac{\langle S \rangle}{M_P} \simeq m_W. 
\eea
Remarkably, the triplet Higgs and the doublet Higgs masses are 
 comparable since exponentially suppressed contributions are negligible. 
Furthermore, the triplet Higgs mass is predicted\footnote{Taking into account 
SUSY breaking, triplet Higgsinos obtain an addtional 
SUSY breaking mass.}, 
\bea
m_3 \simeq \frac{2}{3}m_2 \sim 67~{\rm GeV}. 
\eea
The proton decay constraints in this case are 
\bea
R^{-1} &<& 6.8 \times 10^{17}~{\rm GeV}~({\rm dim}~5), \\
R^{-1} &<& 3.5 \times 10^{17}~{\rm GeV}~({\rm triplet~Higgs~exchange}), \\
R^{-1} &>& 5.5 \times 10^{11}~{\rm GeV}~({\rm X,Y~gauge~boson~exchange}).
%M_5 &<& 8.1 \times 10^{17}~{\rm GeV}~({\rm dim}~5), \\
%M_5 &<& 6.0 \times 10^{17}~{\rm GeV}~({\rm triplet~Higgs~exchange}), \\
%M_5 &>& 10^{16}~{\rm GeV}~({\rm X,Y~gauge~boson~exchange}). 
\eea
The allowed parameters are obtained as 
\bea
%1.0 \times 10^{16}~{\rm GeV} < &M_5& < 6.0 \times 10^{17}~{\rm GeV}, \\
5.5 \times 10^{11}~{\rm GeV} < &R^{-1}& < 3.5 \times 10^{17}~{\rm GeV}, \\
160~{\rm GeV} < &\langle S \rangle& < 8.0~{\rm TeV}, 
\eea
where as in the case with $S$ localized at $y = \pi R$, 
a large $\langle S \rangle$ coressponds to a small $M_5$, 
namely a small $R^{-1}$ and vice versa. 
Thus, the bulk constant $S$ case is very interesting in that 
 the extremely light Higgs triplets is consistent with the proton decay 
 constraints and might provide an alternative signature of GUT 
 in future collider experiments, as discussed in \cite{HabaMaru}. 
%To recover gauge coupling unification, we have only to set 
% the compactification scale $\frac{1}{R}$ to be more than 
% the GUT scale $10^{16}$ GeV and to introduce the brane localized 
% doublets with the triple Higggs mass at $y=0$. 
The detailed analysis of collider signatures of light Higgs triplets 
 is recently presented in Ref.~\cite{Cho}. 
Before summarizing this paper, 
we briefly comment on the gauge coupling unification. 
As discussed in \cite{HabaMaru}, we can consider the possibility that 
extra fields ${\bf 5} + {\bf 5}^*$ are introduced to recover 
the gauge coupling unification. 
However, in the present case, 
since SU(5) is broken at the cutoff scale $M_5$, 
we do not consider the gauge coupling unification.

In summary, 
 we have presented a simple higher dimensional mecahnism of 
 the doublet-triplet splitting in a five dimensional SUSY SU(5) GUT 
 on $S^1/Z_2$. 
The splitting of multiplets is realized 
 by a mass term of Higgs hypermultiplet 
 which explicitly breaks SU(5) gauge symmetry. 
Depending on the sign of mass, 
 zero mode Higgs doublet and triplet are localized on 
 the opposite side of the fixed points. 
The mass splitting is realized due to the differnce of magnitudes 
 of the overlap with a brane localized singlet field 
 not due to the boundary conditions. 
An unnatural fine-tuning of parameters is not necessary. 
As well as a conventional doublet-triplet splitting solution, 
 it has been shown that the weak scale Higgs triplet 
 is consistent with the proton stability. 
This result might provide an alternative signature of GUT 
 in future collider experiments.

We would like to stress that 
the proton stability is ensured by locality without symmetries! 
In a recent orbifold GUT literature \cite{HN}, 
dimension-5 baryon- and lepton-number violating operators are 
forbidden by a $U(1)_R$ symmetry. 
However, dimension-5 and -6 baryon- and lepton-number violating operators 
in our model are strongly suppressed by a small overlap of wave functions 
in spite of an orbifold setup is adopted. 

\vspace*{2cm}

%%%%%%%%%%%%%%%%%%%%%%%%%%%%%%%%%%%%%%%%%%%%%%%%%%%%%%%%%%%%%%%%%%%%%%%%%%%%%%%
%                             Acknowlegdments                                 %
%%%%%%%%%%%%%%%%%%%%%%%%%%%%%%%%%%%%%%%%%%%%%%%%%%%%%%%%%%%%%%%%%%%%%%%%%%%%%%%

\begin{center}
{\bf Acknowledgments}
\end{center}
N.M. thanks R. Kitano and M. Quiros for valuable discussions. 
N.H. is supported by the Grant-in-Aid for Scientific Research, 
Ministry of Education, Science and Culture, Japan 
 (No.14039207, 14046208, 14740164) and 
N.M. is supported 
by Special Postdoctoral Researchers Program at RIKEN.

\vspace{1cm}
%%%%%%%%%%%%%%%%%%%%%%%%%%%%%%%%%%%%%%%%%%%%%%%%%%%%%%%%%%%%%%%%%%%%%%%%%%%%%%%
%                                  References                                 %
%%%%%%%%%%%%%%%%%%%%%%%%%%%%%%%%%%%%%%%%%%%%%%%%%%%%%%%%%%%%%%%%%%%%%%%%%%%%%%%

\end{document}